\documentclass[reprint,twocolumn,aps,pra,nofootinbib,superscriptaddress,amsmath,amssymb,
longbibliography]{revtex4-2}
\usepackage{graphicx}
\usepackage{txfonts,comment}
\usepackage{bm}
\usepackage[colorlinks=true,	linkcolor=blue,urlcolor=blue,anchorcolor=blue,citecolor=blue,bookmarksnumbered]{hyperref}
\usepackage{float}
\usepackage[mathscr]{euscript}
\usepackage[english]{babel}
\usepackage{natbib}
\usepackage{physics}

\def\be{\begin{equation}}
	\def\ee{\end{equation}}
\def\bea{\begin{eqnarray}}
	\def\eea{\end{eqnarray}}
\def\bes{\begin{subequations}}
	\def\ees{\end{subequations}}

\begin{document}
\title{Many-Body Physics with Rydberg Atoms: Quantum Simulation and Non-equilibrium Dynamics}

\author{Zhengyang Bai}
\email{zhybai@nju.edu.cn}
\affiliation{National Laboratory of Solid State Microstructures and School of Physics, Collaborative Innovation Center of Advanced Microstructures, Nanjing University, Nanjing 210093, China}
\author{Cheng Chen}
\email{cheng.chen@iphy.ac.cn}
\affiliation{Institute of Physics, Chinese Academy of Sciences, Beijing 100190, China}

\author{Fan Yang}
\email{fanyangphys@gmail.com}
\affiliation{School of Physics and Zhejiang Key Laboratory of Micro-nano Quantum Chips and Quantum Control, Zhejiang University, Hangzhou 310027, China}
\affiliation{Niels Bohr International Academy, Niels Bohr Institute, University of Copenhagen, DK-2100 Copenhagen, Denmark}

\author{Weibin Li}
\email{weibin.li@nottingham.ac.uk}
\affiliation{School of Physics and Astronomy, and Centre for the Mathematics and Theoretical Physics of Quantum Non-equilibrium Systems, University of Nottingham, Nottingham, G7 2RD, UK}

%\date{\today}	
\begin{abstract}
	    Rydberg atoms, characterized by their strong and long-range dipole-dipole interactions, provide a versatile platform for exploring intriguing collective and many-body effects. Recently, the experimental realization of these effects in dense ensembles and reconfigurable atomic arrays has attracted significant interest, particularly for applications in quantum simulations and non-equilibrium physics. 
        This review focuses on such recent development, discussing the theoretical foundations of the interactions between Rydberg atoms and the ensuing many-body physics, while providing a critical survey of experimental techniques for their precise manipulation and observation. We further discuss recent breakthroughs in leveraging Rydberg collective effects to probe novel many-body phases and non-equilibrium dynamics of these systems. By synthesizing theoretical insights with experimental milestones, we provide a comprehensive perspective on this rapidly evolving field and its transformative potential for future quantum technologies.
\end{abstract}
\maketitle

\section{Introduction}
Rydberg atom systems have become a central platform for quantum many-body studies due to their unique properties. Rydberg atoms combine strong, tunable long-range interactions with high-fidelity control of their electronic states and spatial geometry \cite{Saffman2010, Browaeys2020, Wu2021}.  In tweezer arrays, this combination enables the direct implementation of many-body spin models, constrained Hamiltonians, and programmable interaction graphs, while in thermal vapors, it allows to investigate driven-dissipative nonlinear dynamics with intrinsic coupling between coherent drive, interactions, and loss \cite{Carr_Nonequilibrium_2013,Wadenpfuhl_2023,Ding_2024}. Such capabilities make Rydberg atom platforms uniquely suited for exploring the quantum simulation of exotic many-body physics and probing emergent collective dynamics.

In recent years, we have witnessed rapid progress in the scale and complexity of quantum many-body systems. This includes, for example, programmable dynamics in 51-atom arrays~\cite{Bernien2017}, correlation growth in synthetic Ising models~\cite{Lienhard2018}, nonequilibrium critical scaling through the quantum Kibble-Zurek mechanism~\cite{Keesling2019}, and two-dimensional (2D) quantum phases of hundreds of atoms~\cite{Ebadi2021,Scholl_2021}. In addition to conventional symmetry-breaking phases, programmable constraints and geometry have enabled the exploration of topological and frustrated physics, including experimental probes of the symmetry-protected topological phase~\cite{Leseleuc2019} and spin-liquid physics~\cite{Semeghini2021}. In parallel, nonequilibrium coherent dynamics in driven arrays have clarified mechanisms of slow thermalization and scar-mediated weak ergodicity breaking \cite{turner2018weak,Bluvstein2021}.

Equally important, Rydberg atom systems have opened a route for probing many-body  semiclassical dynamics in open settings. Driven-dissipative ensembles exhibit interaction-induced nonlinearities, optical bistability, hysteresis, limit cycles, synchronization, and ergodicity-breaking behavior~ \cite{Carr_Nonequilibrium_2013,Wadenpfuhl_2023,Ding_2024,YL}. These phenomena are naturally interpreted using mean-field and stability analyses (fixed points, bifurcations, and collective modes), but are rooted in microscopic dipolar or van der Waals (vdW) interactions. This duality between quantum Hamiltonian engineering and nonlinear nonequilibrium dynamics is a key conceptual theme for current Rydberg research, and may pave new routes for quantum sensing applications~\cite{ding2022enhanced}.

 Therefore, the motivation for this review is twofold.  We aim to provide a unified account of many-body quantum simulations in Rydberg atom arrays. On the other hand, it covers  the emerging landscape of semiclassical collective dynamics in dissipative Rydberg atom ensembles. The central question is not only what phases can be engineered, but also how the interaction, driving, geometry, and dissipation jointly determine the real-time pathways to ordering, thermalization, synchronization, and sensing-relevant response. We are aware that this field is evolving rapidly, and new results are emerging daily. Some studies relate to the many-body nature of Rydberg atom settings, but they can be explored and understood from different angles. Hence, we cannot cover every aspect of this field. Readers may find some of them in reviews published parallel to this one. 

 This article is organized into two complementary sections. Section II focuses on  quantum simulation with Rydberg atom arrays (Ising- and XY-type models, constrained dynamics, topological/optimization-oriented settings, and engineered interactions). Section III focuses on nonequilibrium phase transitions in driven-dissipative ensembles, emphasizing the collective phase structure, optical bistability, nonstationary oscillatory states, and ergodicity breaking. By treating these domains within one framework, this review aims to show how quantum many-body dynamics and collective dynamics relate to each other and develop in diverse directions in state-of-the-art Rydberg atom platforms.

	\section{Quantum Simulation of many-body models in Rydberg atom arrays}
	Rydberg atom arrays have emerged as one of the most promising platforms for quantum simulation and computation \cite{Saffman2010, Browaeys2020, Wu2021}.
	The versatility of Rydberg atom arrays stems from a range of control mechanisms, including programmable lattice geometries, site-resolved addressing, and the ability to tune the interactions via laser beams or microwave fields. 
	These capabilities facilitate the exploration of many different types of quantum many-body models, spanning both equilibrium phases and non-equilibrium dynamics. 
	In this section, we review several experimentally feasible quantum models recently realized with Rydberg atom arrays, discussing the relevant technological progress as well as the emergent phenomena of complex quantum matter within these systems.

	\subsection{van der Waals interaction and quantum Ising model}
	The quantum Ising model is one of the most fundamental spin models in condensed matter physics. In the context of Rydberg atom arrays, this model is physically realized by mapping the internal electronic states of the atoms onto effective spin-1/2 particles.
	In this mapping, the stable ground state $\ket{g}$ represents the spin-down state $\ket{\downarrow}$, whereas a highly excited Rydberg state $\ket{r}$ represents the spin-up state $\ket{\uparrow}$ \cite{Browaeys2020, Wu2021}. The dynamics of this system is governed by the transverse-field Ising Hamiltonian~\cite{labuhn2016tunable} (the reduced Planck constant is set as $\hbar\equiv1$ throughout this paper),
	
	\begin{equation}\label{eq:ising_hamiltonian}
	\hat{H} = \sum_i \frac{\Omega}{2}\hat{\sigma}_x^i - \sum_i \Delta \hat{n}_i + \sum_{i<j} V_{ij} \hat{n}_i \hat{n}_j,
	\end{equation}
where $\Omega$ is the Rabi frequency, $\Delta$ is the laser detuning, and $V_{i j}={C_6}/R_{i j}^6$ is the strength of the van der Waals interaction. The tunability of the parameters in the model allows to probe quantum many-body phases of the model~\cite{labuhn2016tunable}. In addition to that, the Rydberg interaction can be strong enough to impose the kinetic constraints in such a quantum Ising model, enabling a number of intriguing phenomena as discussed below.

\begin{figure*}
	\centering
	\includegraphics[width=1\textwidth]{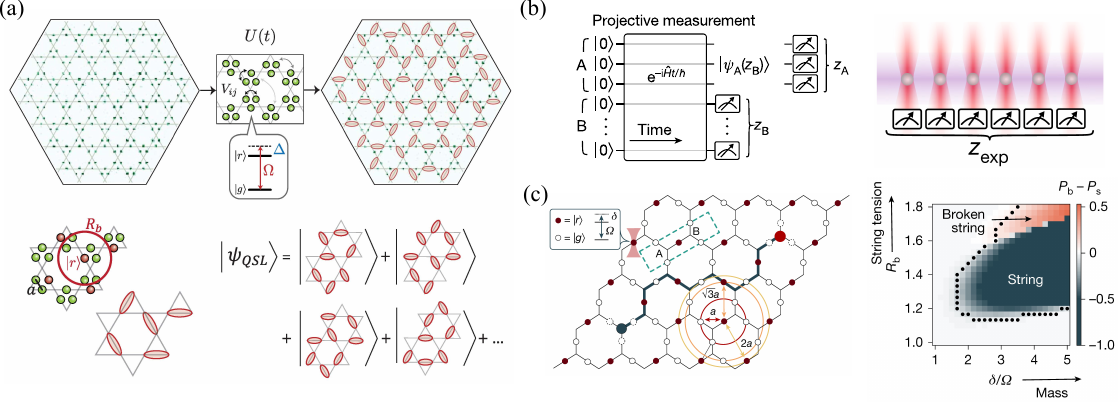}\\
	\caption{Quantum simulation with Ising-type interactions in Rydberg arrays. (a) Realization of a quantum spin liquid on a Kagome lattice, where the dimer constraint is enabled by the Rydberg blockade. (b) Preparing a random quantum state using quantum thermalization and site-resolved imaging in a 1D Rydberg array. (c) Observation of string-breaking in a U(1) lattice gauge theory in (2+1)D. (Reproduced with permission and adapted from Ref.~\cite{Semeghini2021}, Ref.~\cite{choi2023preparing}, and Ref.~\cite{GonzalezCuadra2025}, licensed under a Creative Commons Attribution 4.0 International license.)}\label{Fig_Ising}
\end{figure*}

\subsubsection{Emergent quantum phase transitions}
One of the most prominent kinetic constraints is the Rydberg blockade \cite{Gaetan2009, Urban2009}, in which the strong interaction between atoms prevents more than one Rydberg excitation within a characteristic blockade radius $R_{\text{b}}=(|C_6/\Omega|)^{1/6}$. In a many-body setting, the corresponding quantum phase can be experimentally prepared through an adiabatic annealing process: the system is initialized in a trivial state $\ket{gg\cdots g}$, followed by an adiabatic ramping of the detuning $\Delta(t)$ from large negative values to large positive values.

Based on this approach, the quantum phases of a one-dimensional (1D) Rydberg Ising chain are realized by Bernien et al. \cite{Bernien2017}. In this work, by programming the lattice spacing $d$ and tuning the ratio $R_{\text{b}}/d$, they demonstrated a programmable quantum simulator using atom-by-atom assembly of 51 neutral atoms. They observed phase transitions into spatially ordered Rydberg crystals that break discrete translational symmetries. The experiment verified the preparation of $\mathbb{Z}_2$ ($\ket{rgrg\cdots r}$), $\mathbb{Z}_3$ ($\ket{rggrgg\cdots r}$), and $\mathbb{Z}_4$ ($\ket{rgggrggg\cdots r}$) states across the phase boundaries. The idea is then generalized to two dimensions by Lienhard et al. \cite{Lienhard2018}, who investigated the spatial and temporal growth of antiferromagnetic correlations in dynamically tuned Ising models. They explored the antiferromagnetic phase in 1D chains, square, and triangular lattices with up to 36 atoms, and observed delays in the buildup of correlations between distant sites. Subsequently, two related studies significantly expanded the scale of the atomic quantum simulators to $\sim200$ qubits. In Ref.~\cite{Scholl_2021}, Scholl et al. utilized programmable arrays of up to 196 atoms to realize a two-dimensional Ising model. They probed the antiferromagnetic ordering in square and triangular lattice geometries by dynamically tuning Hamiltonian parameters. Ebadi et al. expanded the simulator to 256 atoms to investigate various quantum phases for different values of the tunable blockade range $R_b/a$ and detuning $\Delta/\Omega$ \cite{Ebadi2021}. They characterized high-fidelity antiferromagnetically ordered states and explored the quantum critical dynamics in (2+1) dimensions.

%\textbf{Z2 quantum spin liquid}\\
Besides the ordered phases demonstrated in the above works, which can be described by the conventional Landau paradigm in terms of spontaneous symmetry breaking, Rydberg arrays further open the possibility to study phase transitions beyond the Landau paradigm. For instance, on a triangular lattice, two distinct ordered phases with $1/3$ and $2/3$ Rydberg filling fractions have been observed \cite{Scholl_2021}. A recent theoretical study suggests the existence of deconfined quantum criticality in this configuration, which describes phase transitions between two ordered phases that break distinct symmetries \cite{bombieri2025deconfined}. The Rydberg blockade can be further used to encode the dimer constraints, which supports a highly nontrivial $Z_2$ quantum spin liquid and a topological phase transition \cite{verresen2021prediction}. The scheme is experimentally demonstrated by Semeghini et al. on a programmable Rydberg array containing 219 atoms \cite{Semeghini2021}. By arranging atoms on the links of a kagome lattice and properly choosing the Rydberg blockade radius, they implemented a frustrated dimer model where each lattice vertex is associated with at most one ``dimer'' represented by a Rydberg excitation [see Fig.~\ref{Fig_Ising}(a)].

    	%\textbf{Quantum Kibble-Zurek mechanism (QKZM)}\\
	In addition to quantum phases, the dynamical preparation approach utilized in this platform opens the door to studying critical dynamics, particularly through the lens of the quantum Kibble-Zurek mechanism (QKZM). For example, Keesling et al. \cite{Keesling2019} used a Rydberg simulator to verify the QKZM associated with quantum phase transitions. They analyzed the growth of spatial correlations when crossing an Ising-type transition at varying speeds, observed scaling universality, and identified corrections beyond the standard QKZM predictions. Zhang et al. \cite{Zhang2025a} explored near-critical Kibble-Zurek scaling in Rydberg arrays when the critical point is smeared by finite-size effects. They demonstrated that precise scaling can be retained in the near-critical regime by appropriately scaling the system size and symmetry-breaking fields. Furthermore, Manovitz et al. \cite{Manovitz2025} experimentally studied the collective coarsening dynamics across a (2+1)-dimensional Ising quantum phase transition. They observed a gradual increase in correlations through the coarsening of antiferromagnetically ordered domains. By preparing seed domains, they demonstrated that the domain boundary curvature drives the coarsening process. They also found that dynamics accelerate as the system approaches the quantum critical point.

	\subsubsection{Quantum thermalization and ergodicity breaking}
	The precise single-site control and readability allows one to probe many-body quench dynamics, such as quantum thermalization and intriguing mechanisms that can break the er~\cite{Bernien2017}. A well-known example is the quantum many-body scar, which can arise in the so-called PXP model. The PXP model in 1D, described by $\hat{H}_{\text{PXP}} = \sum_i \hat{P}_{i-1} \hat{\sigma}_x^i \hat{P}_{i+1}$, with $\hat{P}_i = |g\rangle_i\langle g|$ being a projector onto the ground state of site $i$, emerges naturally from the Rydberg Ising Hamiltonian in the nearest-neighbor blockade regime $V_{i,i+1} \gg \Omega$, where a spin flip can only occur if both neighbors are in the ground state. This model leads to a reduced Hilbert space whose dimension grows in the Fibonacci sequence. The spectrum of the model exhibits a rather counterintuitive phenomenon: while most of the eigenstates obey the eigenstate thermalization hypothesis (ETH), there exists a small set of eigenstates showing atypically large overlaps with the $\mathbb{Z}_2$ ordered state and small entanglement entropies \cite{turner2018weak}. These special eigenstates are called quantum many-body scars (QMBS), which are formed due to a hidden su(2)-like algebra.
    
    Systems supporting QMBS can weakly break ergodicity: while most initial states exhibit rapid information scrambling due to the ETH, slow thermalization can be achieved by preparing an initial state with a large overlap with the QMBS. In Rydberg arrays, the rapid thermalization from a trivial paramagnetic state $\ket{gg\cdots g}$ is studied in Ref.~\cite{kim2018detailed} and characterized in terms of detailed balance \cite{ates2012thermalization}. Recently, Choi et al. show that the thermalization process can also be used to prepare Haar-random state ensembles in the constrained Hilbert space \cite{choi2023preparing} via site-resolved projective measurements [see Fig.~\ref{Fig_Ising}(b)]. Thermalization leads to fast entanglement growth beyond the capability of exact classical simulations, as demonstrated by Shaw et al. \cite{Shaw2024}.
    
    In contrast, the nonthermal behavior from an antiferromagnetic $\mathbb{Z}_2$ state is first observed in Ref.~\cite{Bernien2017}. Here, the initial state has a large overlap with the QMBS featuring equidistantly distributed eigenenergies, leading to a long-time coherent revival. The study of QMBS is subsequently generalized to two dimensions by Bluvstein et al. \cite{Bluvstein2021}, where the scarring dynamics can be controlled by using different lattice geometries. They further apply periodic drives and demonstrate the discrete time crystalline order. In the PXP model, effective time reversal can be easily realized by flipping the sign of the Rabi driving. This facilitates the measurement of the out-of-time-ordered correlator (OTOC), as recently demonstrated by Liang et al. \cite{Liang2025} and Xiang et al. \cite{Xiang2024}. Ref.~\cite{Liang2025} characterizes information spreading by the OTOC and observe persistent oscillations inside a suppressed linear light cone for the initial N\'{e}el state. Ref.~\cite{Xiang2024} explored the collapse-and-revival of quantum information in terms of the OTOC and Holevo information, demonstrating the information backflow.

    Observations of other types of kinetic constraints and ergodicity breaking paradigms have been reported as well. For example, by working in the large-detuning regime $|\Delta|\gg\Omega$, a Heisenberg-type spin model is realized in Ref.~\cite{Kim2024}, where the large anisotropy leads to long-range magnon bound states and constrained spin flip-flops. Such a constrained spin model exhibits strong Hilbert space fragmentation \cite{yang2025probing}, which strongly breaks the ETH by forming exponentially many disconnected Krylov subspaces. In addition, by working in the antiblockade regime, where the detuning $\Delta$ can compensate for the van der Waals interaction shift $V_{ij}$, various distinct kinetic constraints can be engineered for the system. The resulting constrained dynamics can give rise to Krylov-restricted thermalization \cite{thermalization_zhao_2025}, statistical localization \cite{datla2025statistical}, collective nucleation phenomena \cite{Chao2025,Osterholz2026} and prethermal dynamics \cite{darbha2025probing}.

    \subsubsection{Lattice gauge theory and high-energy physics}
    Gauge symmetry is essential for unifying the  fundamental forces in the modern standard model. The lattice version of the gauge theory is formulated on a discretized spacetime lattice, providing a viable way to perform first-principle numerical simulations of the related problems, especially non-perturbative regimes of the quantum electrodynamics (QED) and quantum chromodynamics (QCD) \cite{banuls2020simulating}. The key to realizing LGTs on quantum simulators lies in the engineering of the local gauge constraint, e.g., the Gauss law that relates the bosonic electric field and fermionic matter field. The strong Rydberg interaction often sets the highest energy scale of the dynamics, providing a feasible way to engineer different gauge constraints \cite{surace2020lattice,homeier2023realistic,cheng2024emergent}. For example, in the 1D case, the nearest-neighbor Rydberg blockade can be directly used to realize a spin-$1/2$ quantum link model (QLM) \cite{surace2020lattice}, where the atomic internal states encode the discretized electric field, while the pseudo matter field is introduced between them.
    
   Such a 1D QLM has been experimentally realized recently \cite{Xiang2025,mark2025observation}. In Ref.~\cite{Xiang2025}, a spatiotemporal control is developed to implement a double quench process, which can freeze out the dynamics after the real-time scattering of meson-like excitations. Ref.~\cite{mark2025observation} studies the massless regime of the model and observe the formation of a ballistic plasma and an unexpected long-time memory effect distinct from QMBS. Simulation of the U(1) LGT in (2+1)D is also recently demonstrated by González-Cuadra et al \cite{GonzalezCuadra2025}. In this work, the authors probe the string breaking dynamics on a Kagome lattice [see Fig.~\ref{Fig_Ising}(c)], where the Rydberg blockade encodes the gauge symmetry, while long-range van der Waals interactions are used to tune the string tension.

	\subsubsection{Quantum optimization}
	Another application of the Rydberg atom array is to solve combinatorial optimization problems. For example, finding the maximum independent set (MIS), i.e., the largest set of vertices that are not connected by an edge in a given graph, can be naturally mapped to the problem of finding the many-body ground state of a Rydberg Ising model \cite{pichler2018quantum}. Here, vertices are represented by a single atom, while an edge is established when the interatomic distance is less than the blockade radius. To prepare the ground state for a given graph, both quantum adiabatic annealing (QAA) protocols or quantum approximate optimization algorithm (QAOA) can be applied. An experimental demonstration of this idea was reported in Ref. ~\cite{Ebadi2022}, where up to 289 qubits are used to solve the MIS problem on hard graphs. This work realizes a closed-loop optimization method to test variational algorithms. The results show that the scaling of the success probability is close to the optimal annealing predictions.

    To go beyond unit-disk graphs restricted by the blockade radius, several schemes have been developed to establish tunable qubit connectivity. For example, Kim et al. demonstrated a vertex-splitting scheme to map complex graphs onto a Rydberg simulator \cite{Kim2022}. They used auxiliary atoms as quantum wires to extend the interaction range. The experimental distributions showed high populations for correct MIS solutions. Based on the same idea, Byun et al. solved the MIS problem for Platonic graphs \cite{Byun2022}. Ref.~\cite{nguyen2023quantum} presents an alternative auxiliary-atom based approach, which encodes each vertex variable in a copy gadget, built from a one-dimensional chains of atoms. By further introducing a crossing gadget and a crossing-with-edge gadget to encode the presence or absence of a link, arbitrary connectivity can be realized. Recently, this approach has been experimentally demonstrated in Refs.~\cite{de2025demonstration} and \cite{PRXQuantum.6.020306}.

% Omran et al. (2019) achieved the generation of GHZ states with up to 20 atoms \cite{Omran2019}, also showcased the platform's versatility by distributing entanglement across the array to create remote Bell pairs between edge atoms.

\begin{figure*}
	\centering
	\includegraphics[width=1\textwidth]{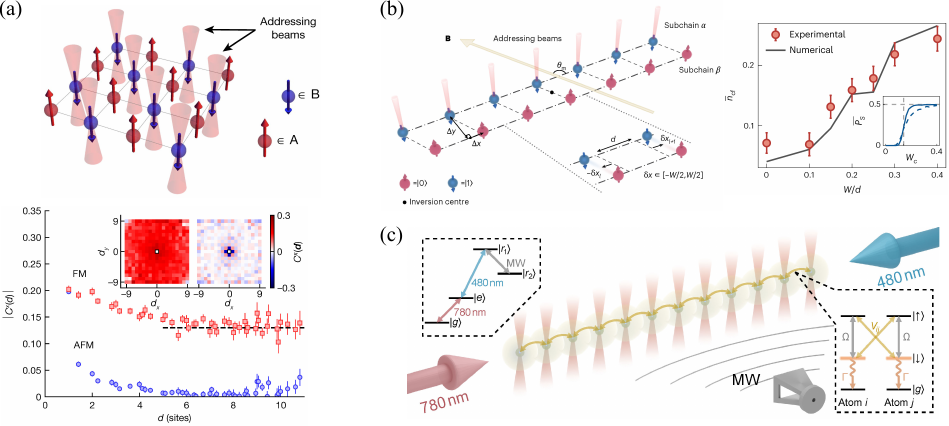}\\
	\caption{Quantum simulation with resonant dipole-dipole interactions in Rydberg arrays. (a) Observation of ferromagnetic and anti-ferromagnetic orders in a dipolar XY model in 2D. (b) Probing average symmetry protected topological order in a Rydberg-atom ladder with programmable disorder. (c) Realization of a non-Hermitian many-body XY model in a 1D Rydberg array. (Reproduced with permission and adapted from Ref.~\cite{Chen2023}, Ref.~\cite{Yue2026}, and Ref.~\cite{Zhang2025}, licensed under a Creative Commons Attribution 4.0 International license.)}\label{Fig_XY}
\end{figure*}

	\subsection{Resonant dipole-dipole interactions and spin XY model}
	The interaction between Rydberg states with opposite parities, such as $\ket{nS}$ and $\ket{nP}$, occurs via resonant dipole-dipole exchange $\propto  {C_3}/{R_{ij}^3} (\hat{\sigma}_i^+ \hat{\sigma}_j^- + \hat{\sigma}_i^- \hat{\sigma}_j^+)$~\cite{Leseleuc2017, Browaeys2020}, where the atomic levels $\ket{nP}$ and $\ket{nS}$ are mapped to spin states $\ket{\uparrow}$ and $\ket{\downarrow}$, respectively. This interaction naturally realizes a long-range spin-$1/2$ XY model as follows:
	\begin{equation}\label{eq:XY_hamiltonian}
		\hat{H}_{XY} = \sum_{i<j} \frac{C_3}{2R_{ij}^3} (\hat{\sigma}_i^x \hat{\sigma}_j^x + \hat{\sigma}_i^y \hat{\sigma}_j^y),
	\end{equation}
	where the coefficient $C_3 \propto (1-3 \cos ^2 \theta_{i j})$ is anisotropic with respect to the angle $\theta$ between the quantization and interatomic axes.

	\subsubsection{Quantum magnetism and exotic many-body phases}
	When the quantization axis is perpendicular to the atomic array plane, the resulting isotropic interactions facilitate the realization of a two-dimensional, long-range dipolar XY model. Recently, Chen \textit{et al.}~\cite{Chen2023} demonstrated the adiabatic preparation of correlated low-temperature states of both the XY ferromagnet and the antiferromagnet [see Fig.~\ref{Fig_XY}(a)]. In the ferromagnetic case, the presence of a long-range XY correlations was reported, a feature indicative of continuous symmetry breaking that is prohibited in the absence of long-range dipolar interaction. Apart from the adiabatic process, in the non-equilibrium dynamics of this model, a spin-squeezed state can be generated.  Bornet  \textit{et al.}~\cite{Bornet2023} demonstrated that quench dynamics from a polarized initial state lead to spin squeezing that is scalable with increasing system size. Furthermore, a quench spectroscopy is used in Ref.~\cite{Chen2025} to characterize the elementary excitations, by monitoring spatial spin correlation dynamics following a quantum quench. Their measurements revealed that ferromagnetic couplings support long-lived spin waves with a nonlinear dispersion relation, whereas the frustrated nature of antiferromagnetic interactions results in a linear dispersion relation. In the study of dipolar XY interactions, recent technological improvements have leveraged the combination of local addressing light shifts and global microwave fields to achieve arbitrary local control within dipolar Rydberg tweezer arrays~\cite{Bornet2024}. This model has also been applied to study the Luttinger liquid behavior of a periodic spin chain~\cite{Emperauger2025}.

	In addition to the conventional quantum magnetism demonstrated in the above studies, recent theoretical studies suggest that the dipolar Rydberg array is a good candidate for $U(1)$ Dirac spin liquid \cite{Bintz2024} and chiral spin liquid \cite{Mao2026, Machado2026}, identifying the system as a potential platform for realizing fractionalized quasi-particles. Based on these predictions, Bornet et al. reported the experimental investigation of a frustrated spin-exchange antiferromagnet using a Rydberg quantum simulator with $N=114$ atoms on a Kagome lattice \cite{Bornet2026}.
    
  The dipolar Rydberg array is highly tunable. For example, in a two-leg ladder, the strong angular dependence of the dipolar interaction allows one to switch off the intrachain hopping, realizing a perfect sublattice symmetry. This facilitates the exploration of the Su–Schrieffer–Heeger (SSH) model and the resulting symmetry-protected topological (SPT) orders, as demonstrated in Ref.~\cite{Leseleuc2019}. In this work, zero-energy edge states as well as their robustness against perturbations were observed. Recently, the model is investigated in a system with programmable atomic position disorder by Yue et al \cite{Yue2026}. The authors observe a disorder-induced, many-body interacting average SPT phase through different disorder realizations [see Fig.~\ref{Fig_XY}(b)]. The dipolar Rydberg system can be further used to study non-Hermitian many-body physics via dissipation engineering. For example, Zhang \textit{et al.} reported the observation of a parity-time symmetry-breaking transition \cite{Zhang2025} in a tunable non-Hermitian XY spin model, where one of the Rydberg states is dressed to a fast-decaying intermediate state [see Fig.~\ref{Fig_XY}(c)].
	
	%\textbf{Floquet engineering and XXZ model)}\\

\subsubsection{Generalizations}
In addition to the dipolar XY model naturally mapped by resonant dipole-dipole interactions, a variety of interesting many-body models can be obtained by engineering Rydberg states. For example, by applying periodic microwave pulse sequences that stroboscopically transfer the population between the spin-up and spin-down states, it is possible to include Ising interactions $\sim\hat{\sigma}_i^z\hat{\sigma}_j^z$. Geier et al. demonstrated such a Floquet engineering of an isolated many-body spin system with disordered Rydberg gas~\cite{Geier2021}, where both a symmetric Heisenberg XXX model and asymmetric XYZ models are realized. The scheme is also successfully implemented in Rydberg array \cite{Scholl2022}, where a tunable XXZ model is engineered, allowing the study of spin diffusion and magnetization dynamics across different regimes of anisotropy.
    
So far, we have introduced models that are time-reversal symmetric. By including more Rydberg levels and applying an external magnetic field, one can break the time-reversal symmetry, for example, by realizing a nontrivial gauge flux for a magnon excitation. The idea is demonstrated by Lienhard et al., who realized a density-dependent Peierls phase through intrinsic spin-orbit coupling of the dipolar exchange interaction system \cite{Lienhard2020}. The authors observed chiral excitation motions and symmetric hole motions of three Rydberg atoms on a triangle, revealing a density-dependent feature of the induced gauge flux. Recent theoretical studies suggest that many-body generalizations of the model support fractional Chern insulators \cite{PRXQuantum.3.030302} and quantum spin liquids \cite{PhysRevResearch.5.013157}. A nontrivial gauge flux can also be included by employing synthetic dimensions in the Rydberg manifold \cite{chen2024strongly,chen2025interaction}, which can exhibit interesting new phenomena under dipolar interactions.

\begin{figure*}
	\centering
	\includegraphics[width=1\textwidth]{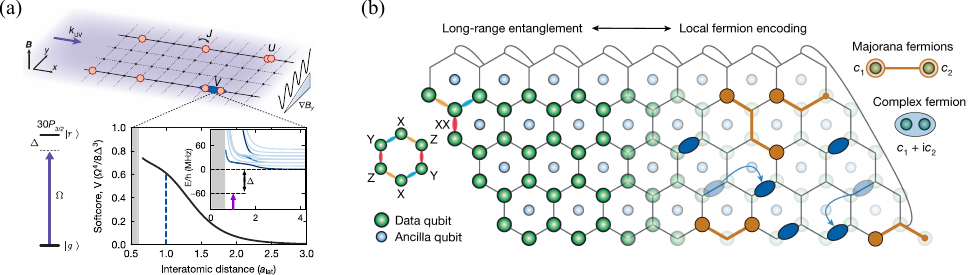}\\
	\caption{Rydberg quantum simulation with interactions induced in the ground-state manifolds. (a) Realization of an extended Bose-Hubbard model in optical lattices, where the long-range density-density interaction is induced by Rydberg dressing. (b) Digital realization of Kitaev honeycomb model in a Rydberg array. (Reproduced with permission and adapted from Ref.~\cite{Weckesser2025} and Ref.~\cite{Evered2025}, licensed under a Creative Commons Attribution 4.0 International license.)}\label{Fig_digital}
\end{figure*}

The original XY model can be further generalized to multi-level regimes, where both direct dipole-dipole and indirect van der Waals interactions coexist. For instance, a $t-J$-type model can be constructed, where spin-up, spin-down particles, and the hole are encoded by three different Rydberg states \cite{Homeier2024}. The scheme is experimentally realized by Qiao et al. in a \cite{Qiao2025a, Qiao2025}. Their implementation of a tunable bosonic $t-J-V$ model allowed for the direct observation of dynamic phase separation and provided evidence for hole pairing mediated by magnetic interactions. In this system, the bosonic $t-J-V$ model is implemented by encoding the effective spin-$1/2$ and hole in three Rydberg states: $|{\downarrow}\rangle = |60S_{1/2}, m_J = 1/2\rangle$ and $|{\uparrow}\rangle = |61S_{1/2}, m_J = 1/2\rangle$, while a $P$-state, $|h\rangle = |60P_{3/2}, m_J = -1/2\rangle$, acts as the hole. In the language of a bosonic $t–J–V$ model, the system is described by Hamiltonian $\hat{H}_{t J V} =\hat{H}_t+\hat{H}_J+\hat{H}_V$ with,
\begin{subequations}\label{eq:tJ_hamiltonian}
\begin{align}
\hat{H}_t & =-\sum_{i<j} \sum_{\sigma=\downarrow, \uparrow} \frac{t_\sigma}{r_{i j}^3}\left(\hat{a}_{i, \sigma}^{\dagger} \hat{a}_{j, h}^{\dagger} \hat{a}_{i, h} \hat{a}_{j, \sigma}+\text { h.c. }\right), \\
\hat{H}_J & =\sum_{i<j} \frac{1}{r_{i j}^6}\left[J^z \hat{S}_i^z \hat{S}_j^z+\frac{J_\perp}{2}\left(\hat{S}_i^{+} \hat{S}_j^{-}+\text {h.c. }\right)\right], \\
\hat{H}_V & =\sum_{i<j} \frac{V}{r_{i j}^6} \hat{n}_i^h \hat{n}_j^h, 
\end{align}
\end{subequations}
where the dipolar tunneling $t_\sigma$ represents the resonant interaction between states of different parity, implementing the hole tunneling (HT). The amplitude $t_\sigma$ is inherently anisotropic, following the dipolar form $t_\sigma \propto (1 - 3\cos^2\theta)$, where $\theta$ is the angle between the interatomic axis and the quantization axis defined by the magnetic field. 
The spin-spin interactions $J$ include the in-plane term $J_\perp$ and the Ising coupling $J_z$. The XY-type coupling $J_\perp$ arises from off-diagonal van der Waals interactions between the $|60S, 61S\rangle$ pair states, resulting in an antiferromagnetic (AFM) exchange. The Ising coupling $J_z$ arises from the diagonal van der Waals interactions between the spin-up and spin-down states.
The diagonal van der Waals interactions between pairs of $|h\rangle$ states lead to a density-density repulsion $V$. Here, the long-range hopping and spin-spin interactions can support exotic phases absent in the conventional local $t-J$ model, e.g., the double supersolid phase in which the crystalline order and two distinct superfluid components coexist, as recently predicted in Ref.~\cite{Chen2025a}.

\subsection{Rydberg dressing and XYZ models}
    Using Rydberg states as media, quantum simulations can be performed within the ground-state manifold, which possesses a much longer lifetime. In this so-called Rydberg dressing scheme, a ground state $|g\rangle$ is typically coupled to a Rydberg state $|r\rangle$ via a laser beam with Rabi frequency $\Omega$ and a large detuning $\Delta$ ($\Delta \gg \Omega$). 
	In the limit of weak coupling, the resulting dressed ground state $| \tilde{g} \rangle \approx |g\rangle + \beta |r\rangle$ acquires a small Rydberg component, where $\beta = \Omega / 2\Delta$ is the admixture coefficient. The atoms remain primarily in the ground state, thereby significantly reducing decoherence from spontaneous emission. They inherit a fraction of the Rydberg-Rydberg interactions. 
	The effective two-body potential $U(r)$ between two dressed atoms at a distance $r$ is derived from the fourth-order perturbation shift, taking the soft-core form \cite{Henkel2010, Johnson2010},
	\begin{equation}\label{eq:Rydbergdressing_hamiltonian}
		U(r) = \frac{\tilde{C}_6}{r^6 + R_c^6}.
	\end{equation}
	Here, $\tilde{C}_6 = \left( \Omega/2\Delta \right)^4 C_6 = \beta^4 C_6$ denotes the effective van der Waals coefficient, where $\beta^4$ represents the probability that both atoms are simultaneously in the Rydberg state during the virtual excitation process.
	The critical dressing radius $R_c = \left( |C_6/2 \Delta| \right)^{1/6}$ defines the spatial extent of the soft core, representing the distance at which the bare Rydberg-Rydberg interaction $V_{rr}(r) = C_6/r^6$ becomes equal to the energy cost of the double excitation $2|\Delta|$.
	The salient feature of $U(r)$ is its behavior in the short-range limit ($r \ll R_c$). As $r \to 0$, the potential approaches a constant plateau value $U(0) = \tilde{C}_6/R_c^6 =  \Omega^4/8\Delta^3$, where the interaction is effectively distance-independent.
    
	The Rydberg-dressed potential provides a versatile tool for quantum simulations. Zeiher et al. report the realization of Rydberg dressing for implementing a 2D synthetic spin lattice \cite{Zeiher2016}. They measure the collective dissipation and coherence after a spin-echo sequence, which was one of the first implementations of dressed interactions in a many-body system. Jau et al. demonstrate the entangling operation between two ground-state atoms in the strong dressing regime \cite{Jau2016}. Subsequently, Zeiher et al. demonstrate coherent many-body dynamics in a 1D Ising spin chain via Rydberg dressing \cite{Zeiher2017}. They engineer a soft-core long-range interaction and observe interaction-driven quantum revivals of magnetization. Similar protocols are then realized in an atomic gas \cite{Borish2020}, where the buildup of correlations and spin dynamics are studied. 
    
    In recent five years, several crucial progresses are reported. For instance, Hollerith et al realize distance-selective interactions by dressing Rydberg macrodimer potentials \cite{Hollerith2022}. Unlike standard vdW interactions, macrodimer dressing creates a potential peak at a specific separation.  Steinert et al demonstrate tunable XYZ models using two-color Rydberg dressing \cite{Steinert2023}, which allows for an independent control of the Ising, flop-flop, and flip-flop terms by adjusting the laser parameters. As a major breakthrough, an extended Bose-Hubbard model is recently realized by the stroboscopic Rydberg dressing \cite{Weckesser2025} [see Fig.~\ref{Fig_digital}(a)], which can substantially increase the coherence time of the system. Based on this technique, the authors probe the correlated dynamics induced by the extended-range repulsively bound pairs. The soft-core, Rydberg dressing induced interactions can also be leveraged to engineer nonclassical states of atom arrays for quantum metrology, e.g., the spin-squeezed states \cite{Eckner2023} and the GHZ states \cite{Cao2024}.

\subsection{Hybrid digital-analog simulations}\label{subsec:digital}
Although the native and induced Rydberg interactions discussed above can be used to perform analog quantum simulations of a variety of intriguing many-body physics, they still restrict the generality of the model that can be simulated. Digital quantum simulation, which decomposes many-body evolution into Trotter sequences that can be realized by a universal logic gate set, has great potential to enable universal quantum simulation \cite{weimer2010rydberg}. The possibility of Rydberg digital simulation is enabled by recent experimental progress in implementing high-fidelity parallel quantum gates between ground-state-encoded atomic qubits \cite{levine2019parallel}. For example, based on high-fidelity quantum gates, Ref.~\cite{Graham2022} executes digital algorithms for solving the quantum chemistry problem and the maximum cut (MaxCut) graph problem. As a major breakthrough, Evered et al demonstrate a two-qubit entangling gate with 99.5\% fidelity on arrays of up to 60 atoms \cite{Evered2023}. The authors utilize fast, single-pulse gates optimized via optimal control theory and suppressed scattering by using atomic dark states. The reported fidelity exceeds the critical error-correction threshold for the surface code.

In a digital Rydberg simulator, the connectivity of the model can also be greatly enhanced, for example, by mid-circuit, coherent transport of ground-state-encoded atoms, as demonstrated in Ref.~\cite{Bluvstein2022}. In this study, by using optical tweezers to move qubits across a 2D space between gate operations, the authors achieved dynamic non-local connectivity, introducing a transformative architecture for neutral-atom processors. This allows for the parallel generation of entangled graph states and the realization of the surface code. They also develop a hybrid analog-digital simulation, which utilizes analog approach to generate entanglement and digital processing to measure it. Based on the same architecture, Evered et al recently report the digital simulation of the Kitaev honeycomb model \cite{Evered2025}. By preparing long-range entangled states with measurement-based schemes, the fermionic statistics are successfully encoded and probed in their qubit realization [see Fig.~\ref{Fig_digital}(b)]. Using a digital Trotterized evolution, they further prepared a non-Abelian spin liquid phase. 

In a more recent work, the hybrid analog-digital scheme is used to simulate a $U(1)$ spin liquid \cite{Geim2026}, where a Floquet driving is used to engineer a tunable analog many-body evolution, while digital operation is employed for non-destructive readout of ground-state-encoded qubits and subsequent qubit refill. In the future, the hybrid analog-digital approach holds promise for significantly broadening the scope of Rydberg quantum simulation, such as the efficient simulation of non-Abelian LGTs \cite{gonzalez2022hardware}, quantum chemistry problems \cite{maskara2025programmable}, and generalized symmetries \cite{warman2025categorical}.

\section{Nonequilibrium Phase Transitions in Rydberg Ensembles}
This section adopts a different tone to examine various novel many-body nonequilibrium phenomena that arise in driven-dissipative Rydberg systems, such as optical bistability, non-stationary states, time crystals, and self-organized criticality. 
Owing to their inherent dissipation and strong interatomic interactions, thermal Rydberg atoms constitute a naturally driven dissipative system. This unique combination makes them a versatile platform for exploring novel many-body non-equilibrium effects in quantum systems.
In the experiment,
atoms are excited from the ground state $|g\rangle$ to the Rydberg state $|r\rangle$ via a two-photon electromagnetically induced transparency (EIT) process~\cite{shaoRydbergSuperatomsArtificial2024a}. This involves a transition through an intermediate state $|e\rangle$, driven by the probe and coupling light fields with Rabi frequencies $\Omega_{p}$ and $\Delta_{c}$, respectively, both of which can be configured as time-dependent parameters. The use of counter-propagating beams in the two-photon excitation scheme through a heated vapor cell ensures that only a narrow, low-velocity class of atoms is excited due to velocity selection effects~\cite{Tanasittikosol_PRA_2012}. The probe transmission is subsequently monitored via a differencing photodetector to measure the transmission difference~\cite{Potvliege_2006, Geier2021}. The EIT-based approach enables continuous, non-destructive, and dynamic monitoring of the transmission of the probe field (the transmission signal is directly proportional to Rydberg population).
To gain a deeper insight into the physical processes discussed here, we begin with a simple two-atom mean-field analysis and obtain the corresponding mean-field phase diagram. Aided by this diagram, we can better comprehend the various nonequilibrium phases that will be addressed in the following sections.
\subsection{Mean field theory}
The dynamics of the Rydberg atom can be described by an effective two-level model~\cite{Lee_Antiferromagnetic_2011, Carr_Nonequilibrium_2013}. We consider an ensemble of $N$ two-level atoms, where the electronic ground state $|g\rangle$ is coupled to excited state $|r\rangle$ by laser fields with an effective Rabi frequency $\Omega$ and detuning $\Delta$. Atoms in state $|r\rangle$ interact
strongly through van der Waals interaction ${V}_{ij}=C_6/|\mathbf{R}_i-\mathbf{R}_j|^6$ with $C_6$ and $\mathbf{R}_{i(j)}$ the dispersion coefficient and location of the $i(j)$-th atom.   The Hamiltonian $\hat{H}$ of the many-body system is given as
$\hat{H}=
\sum_{i} \left[-\Delta \hat{n}_i+\Omega\hat{\sigma}_i^x\right] +\sum_{i<j}{V}_{ij}\hat{n}_i\hat{n}_j$,
% Due to the decay (with rate $\gamma$) from the excited state $|r\rangle$ to ground state $|g\rangle$,
where $\hat{\sigma}_i^x=(|r_i\rangle\langle g_i|+|g_i\rangle\langle r_i|)/2$ flips the atomic state and $\hat{n}_i=|r_i\rangle\langle r_i|$ is the projection operator of the excited state. 
Including the dissipation, dynamics of the system density matrix $\rho$ is modelled by a Lindblad master equation,
\begin{eqnarray}\label{master}
\dot{\rho}(t)={\cal L}\rho(t),
\end{eqnarray}
where ${\cal L}(\cdot)=-i[\hat{H},(\cdot)]+\gamma\sum_{j}( J_j(\cdot)J_j^\dagger-\frac{1}{2}\{J_j^\dagger J_j,(\cdot)\})$ with $J_j=|g_j\rangle\langle r_j|$ is the jump operator~\cite{lee2012collective, Carr_Nonequilibrium_2013}. Here, $\gamma$ denotes the effective decay rate, which accounts for both the finite lifetime of the Rydberg states and dephasing induced by thermal collisions between atoms~\cite{weller2019Interplay,bai2020SelfInduced}. 

For small systems (i.e., about 10 atoms), the quantum master equation can be solved numerically.
However, the Hilbert space of the Hamiltonian grows with $2^N$, while the dimension of the density matrix is $2^{2N}$. The computational complexity prevents us from numerically solving the many-body problem when $N > 10$ with typical computers. Due to the dissipation, many-body correlations may be weak, such that we could employ approximations, such as the mean-field theory and truncated discrete Wigner method~\cite{Schachenmayer_DTWA_2015, singh2022DrivenDissipative, hosseinabadi2025UserFriendly} to simulate the dynamics.

In the mean-field approach, the many-body density matrix $\rho$ is decoupled into tensor products of individual ones, $\hat{\rho} \approx \Pi_i$ $\hat{\rho}_i$. This decoupling essentially ignores correlations between different sites~\cite{diehl2010dynamical}.
This is a good approximation in a three-dimensional system with large number of atoms (typically $10^3\sim10^4$ Rydberg atoms are prepared in the experiment).
%We now introduce the other two spin operators $\hat{\sigma}^y_j=(|g_j\rangle\langle r_j|-|r_j\rangle\langle g_j|)/(2i)$ and $\hat{\sigma}^{z}_j=(|g_j\rangle\langle g_j|-|r_j\rangle\langle r_j|)/2$.
In the mean-field calculation, mean values of spin operators $s^{\mu}_j=\langle\hat{\sigma}^{\mu}_j\rangle~(\mu=x, y, z)$ are calculated, whose dynamics is governed by the following equations of motion,
\begin{subequations}\label{spinmeanfield}
\begin{eqnarray}
&&\frac{ds^x_j}{dt}=-\Delta s^y_j-\frac{\gamma}{2} s^x_j+\sum_{j<k}{V}_{jk}s^y_jn_{r}^k,\\
&&\frac{ds^y_j}{dt}=\Delta s^x_j-\frac{\gamma}{2} s^y_j-\Omega s^z_j-\sum_{j<k}{V}_{jk}s^x_jn_{r}^k,\\
&&\frac{ds^z_j}{dt}=\Omega s^y_j+\frac{\gamma}{2}(1-2s^z_j),
\end{eqnarray}
\end{subequations}
where $n_{r}^j=0.5-s^z_j$ represent the Rydberg population on site $\mathbf{R}_j$. 

\begin{figure}
	\centering
	\includegraphics[width=\linewidth]{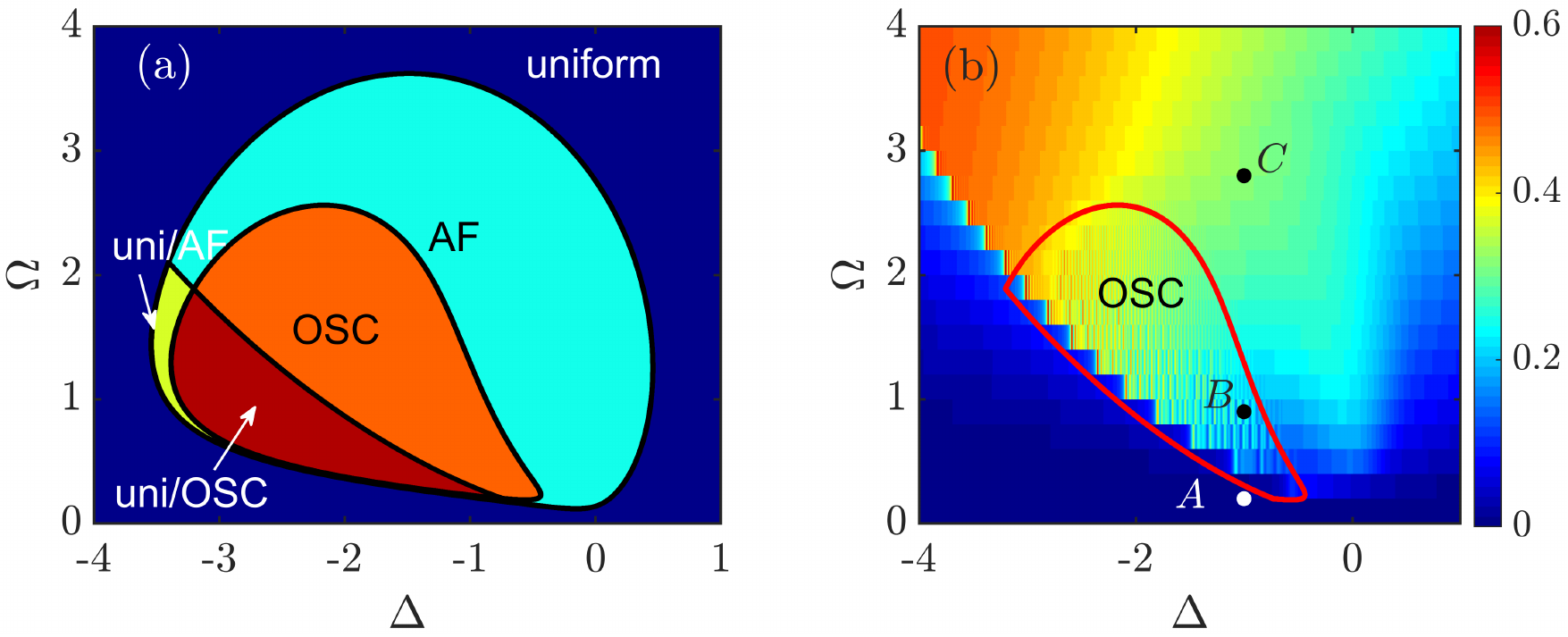}\\
	\caption{\footnotesize {\rm Mean-field phase diagrams.} (a) The phase diagram is mainly occupied by the UNI, AF, and OSC phase.  (b) The many-body phases could be detected by sweeping detuning $\Delta$ from red to blue side linearly (i.e., $\Delta(t)=\Delta_0+a\times t$). The Rydberg population $n_r(t)$  in the $\Delta-\Omega$ plane is sketched. The oscillatory phase can be identified by Rydberg excitation. In the calculation, we set $V_{AB}=-8$, $\gamma=0.5$ and $a=0.01$.}\label{MF_Phase_diagram}
\end{figure}
\begin{figure*}
	\centering
	\includegraphics[width=1\textwidth]{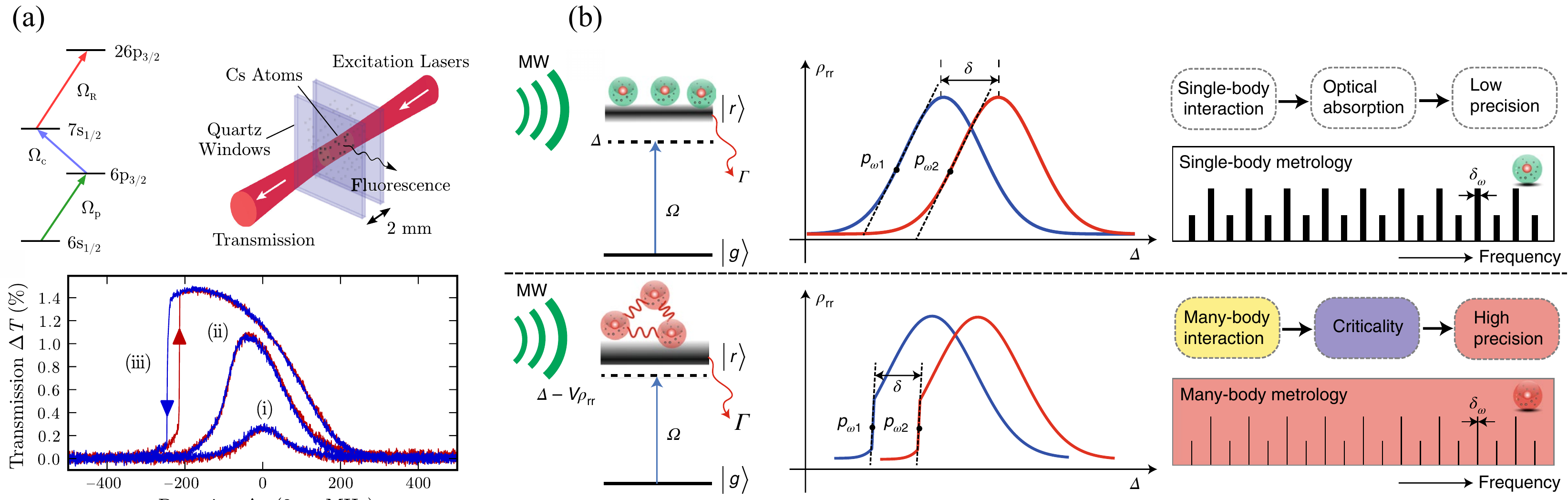}\\
	\caption{\footnotesize (a) A non-equilibrium phase transition is observed in a thermal Rydberg vapor via a multi-photon excitation scheme. (b) Many-body criticality-enhanced metrology for sensing external microwave electric fields in a non-equilibrium Rydberg atomic gas.
   % Figures are adapted from Ref.~\cite{carr2013nonequilibrium} and Ref.~\cite{ding2022enhanced}, respectively.
     (Reproduced with permission and adapted from Ref.~\cite{Carr_Nonequilibrium_2013} and Ref.~\cite{ding2022enhanced}, licensed under a Creative Commons Attribution 4.0 International license.)}\label{bistability_exp}
\end{figure*}

The non-equilibrium phase behavior can be effectively revealed using mean-field theory. For simplicity, we first investigate the minimal model of the quantum many-body system: a bipartite subspace consisting of two sites, labeled $A$ and $B$, which serves as the foundational framework for our analysis~\cite{Lee_Antiferromagnetic_2011,heSuperradianceinducedMultistabilityOnedimensional2022a}. In this case, Eq.~(\ref{spinmeanfield}) is simplified to the following coupled equations of the $A-B$ subspace,
\begin{subequations}\label{eq:AB}
\begin{eqnarray}
&&\frac{ds^x_A}{dt}=-\Delta s^y_A-\frac{\gamma}{2} s^x_A+{V}_{AB}s^y_A n_{r}^B,
\\
&&\frac{ds^y_A}{dt}=\Delta s^x_A-\frac{\gamma}{2} s^y_A-\Omega s^z_A-{V}_{AB}s^x_An_{r}^B,\\
&&\frac{ds^z_A}{dt}=\Omega s^y_A+\frac{\gamma}{2}(1-2s^z_A).
\end{eqnarray}
\end{subequations}
Equations for $B$ site can be obtained by swapping index $A$ and $B$ in Eq.~(\ref{eq:AB}). The fixed points of Eqs.~(\ref{eq:AB}) are found by setting $\dot{s}^x_{A(B)}=\dot{s}^y_{A(B)}=\dot{s}^z_{A(B)}=0$. Two types of fixed points are found: a uniform phase (i.e., $s_z^A=s_z^B$) correspond to spatially homogeneous excitation, and a nonuniform one (i.e., $s_z^A\neq s_z^B$) could contribute to both the antiferromagnetic  and oscillatory phases.
We further analyze the linear stability of the fixed points by calculating eigenvalues $\lambda_j$ of the Jacobian matrix of Eqns. (\ref{eq:AB})~\cite{Strogatz2015Nonlinear}. If the real parts of all eigenvalues are negative, the corresponding solution is stable; otherwise, it is unstable. Mean-field phase diagrams of the bipartite subspace are shown in Fig.~\ref{MF_Phase_diagram}(a). We find that the system is dominated by three cases: uniform (UNI), antiferromagnetic (AF), and oscillatory (OSC). The latter two correspond to nonuniform distribution.

From an experimental perspective, the Rabi frequency $\Omega$ and laser detuning $\Delta$ are the most readily adjustable parameters. To map the predicted mean-field phase diagram, we typically employ a dynamical sweep of either $\Delta(t)$ or $\Omega(t)$. Specifically, the many-body mean-field phases can be probed by linearly sweeping the detuning $\Delta$ from the red-detuned to the blue-detuned side (i.e., $\Delta(t)=\Delta_0+a \times t$). As $\Delta$ approaches resonance, the Rydberg atom population bifurcates from the uniform (UNI) phase into an oscillatory phase. This oscillatory behavior emerges when $\Omega > 0.4$. Notably, while the population dynamics remain stationary in both the uniform and antiferromagnetic (AF) phases, the oscillatory phase is characterized by non-stationary behavior, representing a sharp contrast in the system's dynamical response.

\subsection{The optical bistability induced by Rydberg-Rydberg interaction}
%

%wade2018terahertzdriven
The mean-field phase diagrams, as illustrated in Fig.~\ref{MF_Phase_diagram}(a), clearly reveal the emergence of rich phase behaviors in driven-dissipative Rydberg systems. These nonequilibrium phases have been observed in recent hot-atom EIT experiments~\cite{Carr_Nonequilibrium_2013}. 
By measuring the transmission spectrum of the probe field, the many-body phase behavior can be clearly manifested~\cite{zhangMicrowavecoupledOpticalBistability2025a}.

Experimental observation of nonequilibrium phase transitions in atomic gases typically employs a resonant two-photon or multiphoton excitation scheme, illustrated in Fig.~\ref{bistability_exp}(a). This scheme exploits velocity-selective excitation in a thermal vapor (e.g., Cs or Rb) to focus on a narrow ensemble of atoms, thereby enabling access to a regime where the Rydberg-mediated mean-field shift surpasses the Doppler width. Such a configuration ensures that collective interactions become the dominant energy scale. The corresponding experimental setup is depicted in Fig.~\ref{bistability_exp}(b). The setup utilizes a compact quartz cell filled with thermal alkali vapor. Rydberg excitations are generated by lasers traversing the cell. Atoms are coherently driven to Rydberg states by excitation lasers propagating along the cell’s longitudinal axis, driving the ensemble into the strongly interacting regime with precise control over Rydberg occupancy.

In the experiment atoms within a narrow range of velocity group are excited, the optical transmission $T$  is linearly proportional to the Rydberg excitation  $n_r = \sum_j \langle \hat{n}_j\rangle$ and provides a direct readout of the Rydberg population~\cite{Tanasittikosol_PRA_2012}. 
Fig.~\ref{bistability_exp}(c) depicts the probe transmission $T$ versus Rydberg detuning. With increasing excitation power, the density-dependent shift distorts the resonance into an asymmetric profile (ii) and eventually triggers a transition to intrinsic optical bistability (iii). This regime is marked by a prominent hysteresis loop, where the system exhibits bimodality—occupying either a low- or high-density phase under identical conditions. This behavior aligns with the AF phase predicted in the theoretical phase diagram [Fig.~\ref{MF_Phase_diagram}(b)].

The emergence of optical bistability is critically sensitive to the Rydberg state population. By applying an external control field—such as microwave or terahertz radiation—to resonantly couple adjacent Rydberg states, one can precisely manipulate the excitation density, thereby achieving robust quantum control over the nonequilibrium phase transition~\cite{wade2018terahertzdriven, ding2022enhanced}. This inherent sensitivity presents a promising pathway for high-precision electrometry. By exploiting the enhanced susceptibility near the many-body critical point, the detection sensitivity for microwave fields can be optimized to as low as $49$ nV cm$^{-1}$ Hz$^{-1/2}$~\cite{ ding2022enhanced}. Such performance underscores the potential of bistable Rydberg systems to serve as exceptionally sensitive quantum sensors~\cite{yuanQuantumSensingMicrowave2023}.

% atomic interaction

% optical cavity 

% %LV动力学

% teraherz phase transition\cite{wade2018terahertzdriven}
% \cite{ding2022Enhanced}

% Ding\cite{ding2019Phase}

% Rydberg ions
\begin{figure*}
	\centering
	\includegraphics[width=1\textwidth]{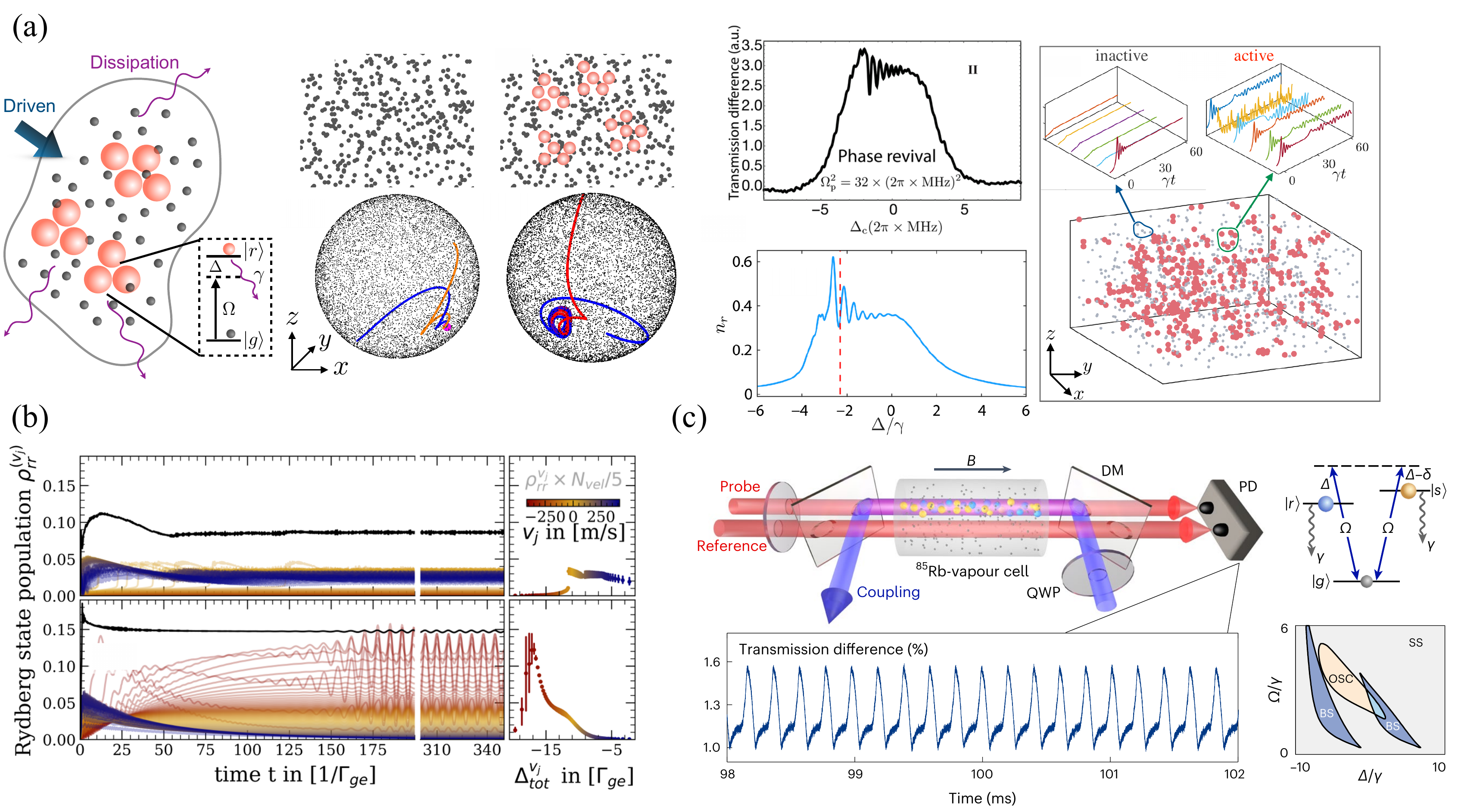}\\
	\caption{\footnotesize {Dissipative continuous time crystal.} (a) Non-stationary phases in a driven-dissipative Rydberg gas. When the interatomic interaction is strong, closely packed Rydberg atoms (red circles) form clusters and become dynamically active, leading to persistent, non-stationary collective dynamics. In phase space, the corresponding trajectories exhibit stable oscillations, forming a limit cycle that signifies Rydberg clustering and synchronized oscillations. (b) Thermal vapor simulation demonstrating the emergence of synchronization. A comparative simulation between uncoupled and coupled velocity classes highlights how synchronization emerges, driven by the Rydberg-density-induced mean field. (c) Realization of a dissipative time crystal in a strongly interacting thermal Rydberg gas. (Reproduced with permission and adapted from Ref.~\cite{Ding_2024}, Ref.~\cite{Wadenpfuhl_2023} and Ref.~\cite{YL}, licensed under a Creative Commons Attribution 4.0 International license.)}\label{oscillation}%Figures are adapted from Ref.~\cite{Ding_2024}, Ref.~\cite{Wadenpfuhl_2023} and Ref.~\cite{YL}, respectively.}\label{oscillation}
\end{figure*}
\subsection{The non-stationary state}

The oscillatory phase induced by limit cycles is illustrated in Fig.~\ref{MF_Phase_diagram}(a). Driven by the interplay between strong Rydberg-Rydberg interactions and dissipation, the system undergoes a Hopf bifurcation, transitioning into a regime of persistent oscillation. This transition is marked by the emergence of interaction-induced dissipative time crystal that spontaneously break continuous time-translation symmetry, leading to exotic non-stationary dynamical phases. Leveraging recent advancements in EIT-based Rydberg detection, which enables continuous, non-destructive monitoring of many-body dynamics, three independent experiments have nearly simultaneously observed these non-stationary phases~\cite{Ding_2024, Wadenpfuhl_2023, YL}.
\subsubsection{Continuous time crystal}
As the Rydberg laser power increases and the system approaches resonance, distinct oscillatory signals emerge in the transmission spectrum. Experimental evidence~\cite{YL} demonstrates a transition from ergodic to ergodicity-breaking dynamics in these systems [see Fig.~\ref{oscillation}(a)]. This broken ergodicity is characterized by persistent, long-lived phase oscillations attributed to the formation of Rydberg excitation clusters within limit-cycle phases. Notably, a significant fraction of the atoms remains dynamically active, exhibiting oscillations over extended durations. Despite initial disparities in amplitude and frequency among individual atoms, the ensemble eventually synchronizes, leading to frequency locking where individual oscillations converge toward a single central frequency.

In a parallel development, Wadenpfuhl et al.~\cite{Wadenpfuhl_2023} reported similar collective synchronization in thermal Rydberg vapors. Despite the inherent dephasing induced by thermal motion, they demonstrated that a global mean-field coupling facilitates frequency and phase entrainment among disparate velocity classes. This mechanism compels the atoms into lockstep oscillations within their respective limit cycles, transforming disordered fluctuations into a coherent, frequency-locked state. As a many-body realization of  Kuramoto model~\cite{Kuramoto2005}, this system—comprising $\sim 10^9$ atoms—offers an ideal, tunable testbed for exploring large-scale synchronization transitions. Fundamentally, the synchronization observed across both velocity classes and spatial coordinates stems from the collective alignment of heterogeneous frequency components. Consequently, these two works share a consistent physical framework, both manifesting the spontaneous locking of diverse individual dynamics.

By applying an external magnetic field, Wu et al.~\cite{YL} realized a spin-1 model using multiple Rydberg states under an external magnetic field. The competition between strong inter-atomic interactions and state-specific shifts triggers a Hopf bifurcation, driving the system into a non-stationary regime. This regime is characterized by limit-cycle dynamics where one Rydberg state is excited at the expense of another. Crucially, the resulting non-decaying oscillations in Rydberg density spontaneously break continuous time-translation symmetry. The robustness of the long-range autocorrelation against temporal noise further confirms the realization of a continuous time crystal (CTC).

\subsubsection{discrete time crystal}
\begin{figure}
	\centering
	\includegraphics[width=\linewidth]{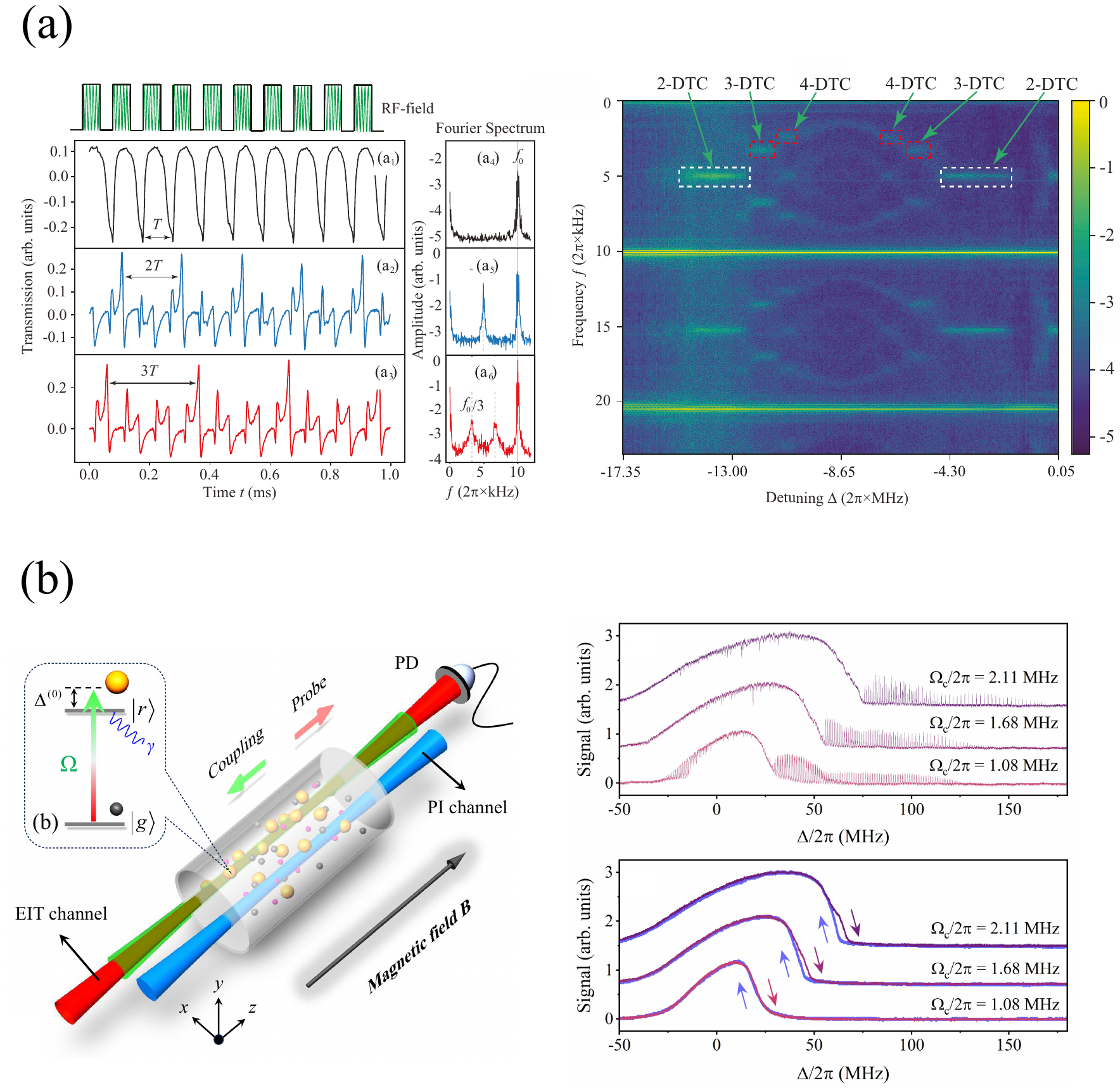}\\
	\caption{\footnotesize (a) Higher-order discrete time crystals are driven by a periodically modulated RF field in Rydberg systems.
(b) Photoionization-induced Floquet driving can generate a robust period-doubling response when the system operates in the bistable region. (Reproduced with permission and adapted from Ref.~\cite{liu2024Higherorder} and Ref.~\cite{jiao2025PhotoionizationInduced}, licensed under a Creative Commons Attribution 4.0 International license.)}\label{DTC}%Figures are adapted from Ref.~\cite{liu2024Higherorder} and Ref.~\cite{jiao2025PhotoionizationInduced}, respectively.}
\end{figure}
Beyond the realization of continuous time crystals via spontaneous Hopf bifurcation, Rydberg systems can host discrete time crystalline (DTC) phases under Floquet driving. These phases emerge from the breaking of discrete time-translation symmetry, typically facilitated by periodic modulation using microwave~\cite{jing2020Atomic} or radio-frequency (RF) electric fields~\cite{Jiao_2017,rotunno2023Detection, liu2022Highly}. Such capabilities underscore the remarkable versatility of Rydberg ensembles as a robust platform for engineering complex many-body dynamics.

In a recent study of discrete time-translation symmetry breaking, Liu et al.\cite{liu2024Higherorder} experimentally reported the emergence of higher-order and fractional discrete time crystals (DTCs) within a Rydberg atomic gas. These phases extend the standard DTC framework through their unique subharmonic responses to a driving period $T$: higher-order DTCs are characterized by a period $nT$ where $n$ is an integer greater than 2, while fractional DTCs represent a non-equilibrium phase where the system’s observables oscillate with a period that is a non-integer rational multiple of $T$ [see Fig.~\ref{DTC}(a)].
By employing a RF electric field to implement periodic Floquet driving, the researchers achieved precise control over the system's dynamics, successfully realizing integer-order DTCs with response periods up to $n=14$ times the driving period. Notably, the study further elucidates the emergence of fractional dynamical behaviors during the transitions between distinct integer DTC phases. 

Beyond external field modulation, Jiao et al. \cite{jiao2024Quantuma, jiao2025PhotoionizationInduced} introduced an alternative driving scheme utilizing a photoionization (PI) laser in conjunction with a magnetic field [see Fig.~\ref{DTC}(b)]. In this configuration, photoelectrons—liberated by the PI laser and guided by the magnetic field—collide with and ionize Rydberg atoms, leading to the formation of a plasma. The resulting plasma-induced Stark shift suppresses further Rydberg excitation until the plasma dissipates, after which the process restarts. This self-sustained cycle establishes a periodic feedback loop that effectively functions as an intrinsic Floquet drive for the Rydberg ensemble.
In a related study, Wang et al.~\cite{Wang_OE_25} investigated the underlying microscopic origins of these nonequilibrium dynamics in thermal Rydberg gases. They experimentally revealed a distinct time-delay effect in the buildup of mean-field interactions, emphasizing the critical role of collision ionization. This temporal delay provides fundamental insights that further elucidate the physical mechanism behind the emergence of such self-sustained oscillations.

By leveraging photoionization-induced Floquet driving, Jiao et al. observed a robust period-doubling response as the system is steered through its bistable region \cite{jiao2025PhotoionizationInduced}. This observation provides a compelling experimental realization of the theoretical framework proposed by Gambetta et al. \cite{gambetta2019Discrete}, which predicts that a periodically driven system with period $T$ can manifest discrete time-translation symmetry breaking within a bistable regime. The underlying dynamics involve driven oscillations between two distinct attractors, a process mediated by an emerging metastable regime. This periodic switching results in the formation of a DTC, fundamentally characterized by persistent subharmonic oscillations at twice the driving period ($2T$).

Recent experiments have unveiled even more intriguing hierarchical dynamics, such as the emergence of a DTC phase driven by the intrinsic periodic oscillations of a CTC \cite{kongkhambut2024Observation, jiao2025Observation}. Fundamentally, these phenomena are manifestations of broken time-related symmetries in non-equilibrium systems. While the majority of current studies focus on the thermodynamic limit, an emerging frontier lies in exploring exotic states within Rydberg optical lattices, such as quantum time crystals  \cite{russo2025Quantumb, xiang2025Quantum} and boundary time crystals \cite{iemini2018Boundary, wang2025Boundary}.  Specifically, in finite systems, fluctuation-driven dynamics lead to unique scaling laws in quantum correlations as a function of system size. These nascent quantum phases represent a fertile ground for discovering new many-body physics.

\begin{figure*}
	\centering
	\includegraphics[width=1\textwidth]{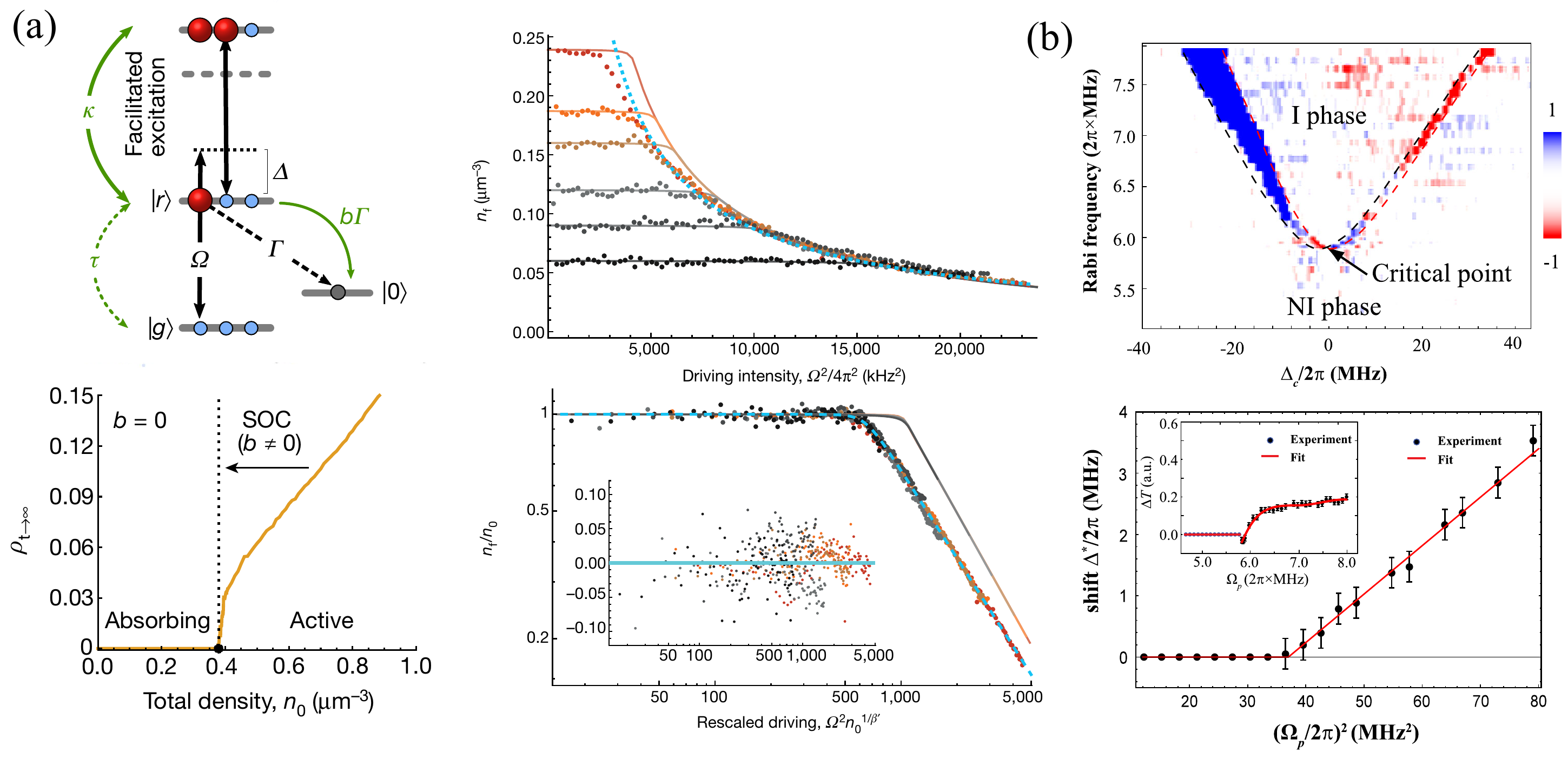}\\
	\caption{\footnotesize (a) Self-organized criticality in a Rydberg atomic gas, showing the scale invariance of the steady state under varying driving intensities. (b) Self-organized dynamics observed in a thermal Rydberg atomic gas. The nonequilibrium phase transition is characterized. (Reproduced with permission and adapted from Ref.~\cite{helmrich2020Signatures} and Ref.~\cite{ding2020Phase}, licensed under a Creative Commons Attribution 4.0 International license.)}\label{SOC}%Figures are adapted from Ref.~\cite{helmrich2020Signatures} and Ref.~\cite{ding2020Phase}, respectively. } \label{SOC}
\end{figure*}
\subsection{The absorbing state phase transition and self-organized criticality}
Beyond the exploration of time crystalline phases and bistability, absorbing state phase transitions under multiplicative noise have garnered significant attention. Near the phase transition point, self-organized criticality (SOC) provides an elegant framework to explain the abundance of scale-invariant systems observed in nature~\cite{bak1987Selforganized}. SOC is widely invoked to explain the emergence of scale-invariant dynamics in diverse complex systems. Its manifestations range from geophysical and ecological events, such as earthquakes~\cite{Earthquakes_1989} and forest fires~\cite{forest-fire_1992}, to biological and technological processes including neural activity~\cite{Neuronal_2012}, disease propagation~\cite{Epidemic_2015}, and the behavior of electrical circuits~\cite{World2000}.

By leveraging the Rydberg-facilitated excitation mechanism, recent experiments have successfully captured hallmark signatures of SOC within Rydberg ensembles \cite{helmrich2020Signatures}. In regimes dominated by strong dephasing, the coherent degrees of freedom are adiabatically eliminated, reducing the many-body dynamics to a set of classical-like rate equations. Under these conditions, the system effectively mirrors the behavior of absorbing state phase transitions typically observed in classical stochastic systems. As demonstrated in Ref.~\cite{helmrich2020Signatures}, a driven-dissipative gas of ultracold atoms can spontaneously evolve into a scale-invariant stationary state, characterized by a universal scaling function and power-law distributed excitation avalanches. This emergent SOC behavior is sustained by a dynamical feedback loop: the intricate interplay between facilitated excitation and spontaneous decay self-steers the ensemble toward a critical attractor, thereby bypassing the requirement for precise fine-tuning of external parameters.

Complementary to the studies in ultracold gases, Ding et al.~\cite{ding2020Phase} demonstrated that self-organizing dynamics and non-equilibrium phase transitions can also be realized in a thermal Rydberg vapor. By utilizing EIT, they mapped out a complex phase diagram where the competition between ground-state excitation and Rydberg-Rydberg interactions leads to self-organized steady states. These findings reveal that even in a high-temperature environment dominated by Doppler broadening, the facilitated excitation mechanism can drive the system toward a critical regime characterized by long-range temporal correlations.

Expanding the observations of SOC in Rydberg gases, Brady et al.~\cite{brady2024Anomalousb} recently elucidated the role of network dynamics in determining the universality class of these non-equilibrium transitions. By employing Monte Carlo simulations and machine learning, they demonstrated that the universality class can be tuned from standard directed percolation in frozen gases to anomalous directed percolation and mean-field behavior in the presence of atomic motion. This transition is driven by the emergence of L\'{e}vy-flight-type long-range excitations as the network becomes dynamic. Crucially, they showed that the SOC process is robust against the inherent decay of Rydberg states, but the specific critical exponents are sensitive to the underlying transport properties of the atoms. This adds a new dimension to the study of Rydberg-based SOC, highlighting the platform's potential to simulate complex spreading processes and tunable non-equilibrium phase transitions beyond the static limit.

The realization of SOC in such highly controllable systems not only bridges the gap between microscopic quantum dynamics and macroscopic non-equilibrium phenomena but also provides a robust framework for understanding the emergence of complexity and many-body correlations in open quantum systems.
A central theme across these studies is the rigorous determination of critical exponents, which define the universality class of Rydberg-based SOC. While experiments in ultracold gases highlighted a distinct departure from mean-field predictions due to many-body correlations~\cite{helmrich2020Signatures}, subsequent research in thermal vapors confirmed the robustness of these power-law signatures even in dissipative environments~\cite{ding2020Phase}. By incorporating atomic motion into Monte Carlo simulations, Brady et al. \cite{brady2024Anomalousb} theoretically demonstrated that the universality class itself is tunable, as evidenced by a continuous shift in critical exponents. %the system can transition from directed percolation to anomalous directed percolation

In contrast to the rate-equation descriptions applicable in high-dephasing limits, moving toward the deep quantum regime reveals a far richer phase landscape. In this regime, as the competition between quantum coherence and classical noise intensifies, the nature of the transition undergoes a fundamental shift. As elucidated by Marcuzzi et al.~\cite{marcuzzi2016Absorbing}, the system can transition from the standard continuous directed percolation class to a first-order (discontinuous) phase transition. This regime is governed by a bicritical point where quantum and classical fluctuations exist on equal terms, giving rise to universal features distinct from classical universality. Within this first-order framework, a slow incoherent pump can trigger the emergence of a CTC through self-organized bistability, accompanied by coherent switching dynamics between the absorbing and active states~\cite{xiang2024Selforganized}. By suppressing dephasing through optical trapping or lattice confinement, Rydberg systems provide a promising platform to realize these uniquely quantum phase transitions, offering a window into many-body physics beyond the reach of classical stochastic models.

%This work significantly extends the scope of Rydberg-based SOC research, suggesting that the inherent robustness of these self-organizing processes allows for the exploration of complex many-body physics in relatively simple, room-temperature experimental platforms.
%
%On the experimental aspect, Jiao et al. ~\cite{jiao2025PhotoionizationInduced} demonstrated the bistable-mediated DTC in a thermal Rydberg ensemble. 

%A spatially separated photoionization (PI) laser (with a wavelength below the photoelectric effect threshold for Cs, 653 nm) is applied to create free electrons from the cell wall with adsorbed Cs atoms [33]. In the presence of a magnetic field along the light propagation direction, these photoelectrons are rotated around the magnetic-field vector and collide with Rydberg atoms created by the EIT excitation. These collisions lead to fast ionization of the Rydberg atoms in less than a few μs [34,35], which form the plasma under a certain condition [29,35]. The plasma creates electric fields that suppress Rydberg excitation via the Stark effect.

%~\cite{liu2024Higherorder}.
% 一些其他的拓展

% \subsection{The influence on electric fields induced by spontaneously ionized Rydberg atoms}

% \subsection{The discussion on optical cavity system}

% \subsection{Cellular Automata}

% \subsection{Tipping point}

% \subsection{AI}

\section{Progresses and outlooks}
The rapid experimental progress of Rydberg system has enabled a variety of new possibilities in studying interesting many-body physics. In this section, we briefly discuss some of these progresses and possibilities.

\textit{Improving the scalability of atom arrays}. To achieve the quantum advantage in simulating many-body physics, a large system size is required, especially for fault-tolerant simulations that can consume large amount of physical qubits. Experimental efforts toward a fast, continuous filling of large atom arrays have now attracted wide attentions. For example, Chiu et al. recently demonstrate the continuous operation of a 3000-qubit system \cite{Chiu2025}. Utilizing optical lattice ``conveyor belts'' for high-rate atom reloading, the researchers maintained a large-scale array for over 2 hours without disrupting the coherence of stored qubits, paving the way for fault-tolerant, deep-circuit evolution. Recently, Lin et al. developed AI-enabled parallel assembly to create defect-free arrays of thousands of neutral atoms \cite{Lin2025}. By using real-time monitoring and high-speed feedback, the system drastically improves the initialization efficiency and scalability of large-scale Rydberg simulators.

\textit{Control of atomic motional states}. While most existing Rydberg simulation protocols only use internal-state degree of freedoms, control of atomic motional states can give rise to a rich variety of many-body phenomena, e.g., complex molecular dynamics \cite{magoni2023molecular}, spin-phonon scattering \cite{magoni2024coherent}, and exotic quantum phases \cite{zhang2025many}. Experimental progresses along this direction have been reported as well. Utilizing an ultrafast excitation scheme, strong spin-motion coupling emerging from the strong van der Waals interaction has been observed in Ref.~\cite{bharti2024strong}. Subsequently, spin-motion coupling emerging from the dipole-dipole interaction has been observed by a Stern-Gerlach-like experiment \cite{gabriel2025probing}. Fully coherent control of atomic motional state is also promising, as recently demonstrated by Lienhard et al. in terms of squeezing of a single-atom motional degree of freedom \cite{lienhard2025generation}, as well as by Shaw et al. who successfully entangle the motional state of two atoms via Rydberg interaction \cite{shaw2025erasure}.
    
\textit{Quantum simulation of fermionic models}. Besides the spin and phonon dynamics, it is highly valuable to study fermionic models that have a wide range of realistic applications, e.g., for simulating properties of electrons in real materials. Optical tweezer array provides a hardware-efficient platform to simulate fermion statistics by loading real fermionic atoms into the trap. Several proof-of-principle experiments already demonstrate the tunneling gate operations between nearby tweezer arrays \cite{spar2022realization,young2022tweezer}. Recent theoretical works show that, when equipped with Rydberg gates \cite{gonzalez2023fermionic}, this platform is promising for programmable and even fault-tolerant fermionic quantum simulators \cite{ott2025error,schuckert2024fault}.

\textit{Error mitigation and fault-tolerant schemes}.
Rydberg-atom based quantum simulation is still in the noise intermediate-scale quantum (NISQ) regime. As an important direction, developing noise reduction or error mitigation schemes holds promise to further improve the performance of the Rydberg quantum simulator. For instance, circular Rydberg states can suppress the error by increasing the coherence time of the simulation \cite{nguyen2018towards,holzl2024long,machu2026nondestructive}. In the case when error already occurs, erasure conversion \cite{wu2022erasure} allows one to detect certain types of the leakage error at a known location, which can be used to purify the simulation outcome based on post-selection, as demonstrated in Refs.~\cite{scholl2023erasure,senoo2026high}.

In the long run, high-fidelity universal logic gates can enable a universal digital quantum simulation, as discussed in Sec.~\ref{subsec:digital}. Following this route, developing error correctable, hardware efficient, and even fault-tolerant schemes are the important next steps. Several key technological ingredients are demonstrated, e.g., individual addressing \cite{Radnaev2025}, mid-circuit measurement \cite{singh2023mid,ma2023high,anand2024dual} and real-time feedforward \cite{lis2023midcircuit}. As a milestone, Bluvstein et al. demonstrate a logical quantum processor operating with up to 280 physical qubits and 40 logical qubits \cite{Bluvstein2024}. They also show that logical encoding can outperform physical qubit fidelities. Based on this result, a fault-tolerant architecture for universal quantum computation is recently demonstrated in Ref.~\cite{Bluvstein2026}, which includes the implementation of transversal gates, lattice surgery, and mid-circuit qubit reuse, paving the way for deep-circuit logical operations.

\textit{Emergent non-equilibrium phases in Rydberg ensembles}.
The rich landscape of non-equilibrium many-body phases in Rydberg ensembles offers fertile ground for further exploration. For instance,
integrating Rydberg atom ensembles into high-finesse optical cavities opens up a fascinating frontier for exploring rich, non-equilibrium many-body dynamics driven by the competition between strong long-range interactions, coherent driving, and dissipation~\cite{Georgakopoulos_2019,cavity_2026, Yasir2026}. By dressing the atomic states into cavity Rydberg polaritons, recent studies have shown that the effective light-atom interaction strength can be significantly boosted~\cite{Wang2023Cavity}. Moreover, the long-range, photon-mediated interactions inherently present in Rydberg ensembles provide an ideal testbed for studying non-Hermitian physics~\cite{Zhang2025, Wang2026NonHermitian}. In these driven-dissipative architectures, the exceptional points and dissipation-induced phase transitions can be dynamically tailored by tuning the dark-state polariton rotation angle or cavity loss rates, offering an unprecedented look into quantum criticality away from thermodynamic equilibrium~\cite{EP_2026}. Beyond fundamental interest, these non-equilibrium phases and their associated phase boundaries hold immense promise for quantum metrology. When a Rydberg time crystal is prepared precisely at the critical edge of a non-equilibrium phase transition, its collective many-body response becomes extraordinarily sensitive to external perturbations. This property can be harnessed for high-precision electrometry~\cite{moon2026Sensing, Liu2026Enhanced,xue2026Enhanced}. Expanding on these collective dynamics, recent research~\cite{liu2026timeseries} utilizes interacting many-body Rydberg systems for time series forecasting, revealing that emergent collective amplification near such phase transitions drastically improves prediction accuracy.

%Furthermore, the dissipative nature of the cavity enables non-destructive, real-time monitoring of the system's interior. By extracting the scaling behavior of the cavity output field and monitoring the saturation of the photon correlation functions, researchers can directly map out universal scaling functions and measure non-equilibrium critical exponents with unprecedented accuracy, paving the way for open-system quantum simulation and sensing~\cite{brady2024Anomalousb}.

\section{Conclusion}
%The systematic classification of interactions in Rydberg atom arrays is essential for mapping appropriate physical models. By sequentially exploring van der Waals, resonant dipole-dipole, and dressed interaction regimes, researchers have established this platform as an exceptionally versatile tool for many-body simulation, optimization, and fault-tolerant quantum computation.

 By exploring van der Waals, resonant dipole-dipole, and dressed interaction, Rydberg atoms become a broad, experimentally accessable setting that spans  quantum simulation of coherent dynamics and dissipative collective phases. In atom arrays, precise control over geometry and interactions has enabled exploration of equilibrium ordering, critical dynamics, constrained Hilbert-space physics, and topological phenomena at scales approaching or exceeding classical tractability \cite{Bernien2017,Keesling2019,Ebadi2021,Scholl_2021,Semeghini2021}. In the thermal atom ensembles, interaction-driven nonlinear response has provide a testbed on bistability, synchronization, and ergodicity-breaking dynamical phases~\cite{Carr_Nonequilibrium_2013,Wadenpfuhl_2023,Ding_2024}. Though seemingly separate, the many-body and classical viewpoints are complementary rather than separate. The mean-field approach provides intuition for observed nonequilibrium phases, while the programmable quantum simulators provide microscopic control to test where semiclassical descriptions fail or require genuine quantum corrections. This interplay is especially relevant near critical regions, where sensitivity enhancement, long-lived oscillations, and slow relaxation make Rydberg atom settings attractive for both fundamental studies and metrological applications.

Looking ahead, there is more exciting physics to explore with the Rydberg atom platform, such as engineering their interactions and dynamics, quantum simulating  many-body models, and exploring algorithm using Rydberg atom gates. These studies are not important for understanding novel physics using the quantum simulator, but also providing new insights into quantum technological applications in sensing and computation.

\begin{acknowledgements}
W. L. acknowledges support
from the EPSRC through Grant No. EP/W015641/1. Z. B. acknowledges support from the National Natural Science Foundation of China under Grant 12274131. C. C. acknowledges the support from Quantum Science and Technology-National Science and Technology Major Project (2024ZD0301700) and the start-up grant from IOP-CAS. F. Y.
acknowledges support from the Firuza Foundation Fellowship.
\end{acknowledgements}
 
%\bibliography{References}


\end{document}